\documentclass[12pt,a4paper,twoside]{article}
\usepackage{epsfig}

\newcommand{\be}{\begin{equation}}
\newcommand{\ee}{\end{equation}}
\newcommand{\beqs}{\begin{eqnarray}}
\newcommand{\eeqs}{\end{eqnarray}}
\newcommand{\LL}{{\cal L}}
\newcommand{\M}{{\bf M}}
\newcommand{\tM}{{\tilde {\bf M}}}
\newcommand{\tr}{{\rm tr}}
\newcommand{\half}{{1 \over 2}}
\newcommand{\m}{{\rm m}}

\def\NP{{\it Nucl. Phys.\ }}
\def\PL{{\it Phys. Lett.\ }}

\def\PR{{\it Phys. Rev.\ }}
\def\PRL{{\it Phys. Rev. Lett.\ }}

\def\JMP{{\it J. Math. Phys.\ }}

\def\MPL{{\it Mod. Phys. Lett. A\ }}

\def\LNC{{\it Lett. Nuovo Cimento \ }}

\expandafter\ifx\csname mathbbm\endcsname\relax

\else

\fi
\begin{document}
\begin{titlepage}
\begin{flushleft}  
       \hfill                      {\tt hep-th/9806189}\\
       \hfill                      UUITP-7/98\\
       \hfill                       June 1998\\
\end{flushleft}
\vspace*{3mm}
\begin{center}
{\LARGE Generalized Calogero-Sutherland systems from many-matrix models
\\}
\vspace*{12mm}
\large Alexios P. Polychronakos\footnote{E-mail:
poly@teorfys.uu.se} \\
\vspace*{5mm}
{\em Institutionen f\"{o}r teoretisk fysik, Box 803 \\
S-751 08  Uppsala, Sweden \/}\\
\vspace*{4mm}
and\\
\vspace*{4mm}
{\em Physics Department, University of Ioannina \\
45110 Ioannina, Greece\/}\\
\vspace*{15mm}
\end{center}

\begin{abstract}
We construct generalizations of the Calogero-Sutherland-Moser
system by appropriately reducing a model involving many 
unitary matrices. The resulting systems consist of particles on
the circle with internal degrees of freedom, coupled through
modifications of the inverse-square potential. The coupling
involves $SU(M)$ non-invariant (anti)ferromagnetic interactions 
of the internal degrees of freedom. The systems are shown to be
integrable and the spectrum and wavefunctions of the quantum 
version are derived.

\end{abstract}

\vspace*{10mm}

\end{titlepage}

The inverse-square interacting particle system \cite{Cal,Suth,Mos} 
and its spin generalizations \cite{GH,Woj,HH,Kaw,MP1,HW} are important models
of many-body systems, due to their exact solvability and
intimate connection to spin chain systems \cite{Hal,Sha,FM,Scal},
2-dimensional Yang-Mills theories \cite{GN,MP2,LSK} etc.

A useful route for studying these systems is through reductions of
appropriate matrix models \cite{KKS,OP}. Although this does not
give rise to all known models in the quantum case, it is still
a very direct way to show integrability and solvability. In this paper,
we will explore this approach further and show how, starting with
many matrices effectively coupled through constraints, we can derive
further generalizations of these systems with internal degrees of freedom.

The starting point will be the many-matrix lagrangian
\be
L = \tr \sum_{i=1}^M -{1\over 2 a_i} ( U_i^{-1} {\dot U}_i )^2
\ee
The $U_i$ are $N \times N$ unitary matrices depending on time $t$
and overdot stands for time derivative. The $a_i$ are real positive 
parameters. For convenience, we shall normalize the $a_i$ such that
\be
\sum_{i=1}^M a_i = 1
\ee

The above system is clearly integrable and solvable, being nothing
more than a collection of independent unitary matrix models. In the absence
of further constraints it would give rise to a collection of
independent spin-generalizations
of the Sutherland model. It is, however, possible to reduce it in a
different way: the above model is invariant under (independent) left-
and right-multiplications of the $U_i$ by constant matrices. This
manifests in the existence of conserved quantities, namely
\be
R_i = -{i\over a_i} U_i^{-1} {\dot U}_i ~,~~~~ 
L_i = {i \over a_i} {\dot U}_i U_i^{-1}
\ee
The above conserved hermitian matrices Poisson-commute with themselves
to the $U(N)$ algebra and mutually to zero. We will choose to
fix the values of the following time-independent quantities:
\be
-{i \over a_i} U_i^{-1} {\dot U}_i + {i \over a_{i+1}}
{\dot U}_{i+1} U_{i+1}^{-1} = P_i
\label{cond}
\ee
where we have adopted periodic conditions in  $i$, that is,
$M+1 \equiv 1$, $0 \equiv M$. Clearly the chosen expressions for the $P_i$,
being the sum of independent $U(N)$ generators,
also constitute mutually commuting Poisson-$U(N)$ matrices.

The key to a successful reduction of this system to particles lies in 
an appropriate parametrization of the $U_i$. We will choose
\be
U_i = W_{i-1}^{-1} \, \Lambda^{a_i} \, W_i
\label{param}
\ee
where $W_i$ are unitary matrices and $\Lambda = diag(e^{i \theta_n})$
is a diagonal unitary matrix. The reason
for the choice of exponents in (\ref{param}) will be apparent in
the sequel. From the relation
\be
U \equiv U_1 \cdots U_N = W_N^{-1} \, \Lambda \, W_N
\label{Uall}
\ee
we see that $W_N$ and $e^{i\theta_n}$ are the diagonalizer and the eigenvalues
of the matrix $U$, and the remaining $W_i$ are determined recursively
from (\ref{param}). The above parametrization has the redundancy
generated by left-multiplication of all $W_i$ by the same diagonal
matrix. This will lead to a `gauge constraint' later on. 

With the above parametrization the expressions for the lagrangian
and the $P_i$ become, after some algebra,
\be
L = \tr \sum_{i=1}^M \half \left\{ -(\Lambda^{-1} {\dot \Lambda})^2
+ \left({1 \over a_i} + {1 \over a_{i+1}}\right) \LL_i^2
-{2 \over a_i} \LL_{i-1} \Lambda^{a_i} \LL_i \Lambda^{-a_i}
\right\}
\label{newL}
\ee
\be
 W_i P_i W_i^{-1} \equiv K_i =
\left({1 \over a_i} + {1 \over a_{i+1}}\right) \LL_i
-{1 \over a_i} \Lambda^{-a_i} \LL_{i-1} \Lambda^{a_i}
-{1 \over a_{i+1}} \Lambda^{a_{i+1}} \LL_{i+1} \Lambda^{-a_{i+1}}
\label{newK}
\ee
where we defined
\be
{\dot W}_i W_i^{-1} = i \LL_i
\ee
{}From the structure of the above lagrangian, we see that the
generator of left-rotations for $W_i$ (found by differentiating
$L$ with respect to $i {\dot W}_i W_i^{-1}$) is exactly $K_i$
as expressed in (\ref{newK}). Therefore, the $K_i$ are also
mutually commuting Poisson-$U(N)$ generators. This justifies the
choice of exponents in (\ref{param}), since any other choice
would spoil the decoupling and $U(N)$ nature of the $K_i$.
We stress, however, that the $K_i$ are no more conserved and are,
thus, dynamical quantities. Further, from (\ref{newK}) we see that
the diagonal elements of $K_i$ must satisfy
\be
\sum_{i=1}^M K_{i nn} = 0 ~~~{\rm for~each}~n
\label{Gauss}
\ee
which is the `Gauss law' originating from the `gauge invariance'
of the parametrization (\ref{param}) as stated previously. The 
off-diagonal elements of $\LL_i$ satisfy
\be
\sum_{j=1}^M \M_{ij} (i\theta_{mn}) \LL_{j mn} = K_{i mn}
\ee
where $\theta_{mn} \equiv \theta_m - \theta_n$ and the matrix $\M (x)$
is defined as
\be
\M_{ij} (x) = \left({1 \over a_i} + {1 \over a_{i+1}}\right) 
\delta_{ij} -{1 \over a_i} e^{- a_i x} \delta_{i,j+1}
-{1 \over a_{i+1}} e^{a_{i+1} x} \delta_{i+1,j}
\ee
(again, taking $\delta_{i,M+1} = \delta_{i,1}$).
In terms of the matrices $K_i$ and the angles $\theta_n$ the
hamiltonian (which is identical to the lagrangian since it only contains
kinetic terms) becomes
\be
H = \half \sum_{n=1}^N p_n^2 + \half \sum_{m\neq n=1}^N
\sum_{i,j=1}^M \M_{ij}^{-1} (i\theta_{mn}) K_{i mn} K_{j nm} 
+ \half \sum_{n=1}^N \sum_{i,j=1}^M \tM_{ij}^{-1} (0) K_{i nn} K_{j nn}
\ee
where $p_n = {\dot \theta}_n$ is the canonical momentum of $\theta_n$
and $\tM (0)$ is the matrix $\M (0)$ projected
to the subspace orthogonal to its zero-eigenvalue eigenvector.

In order to invert the matrix $\M$, we note that it is essentially
a discrete second derivative on a set of points at distances $a_i$,
conjugated with an exponential factor. Generalizing from the case 
of equally-spaced points (that is, the Cartan matrix of $SU(M)$),
where the inverse is proportional to the distance $|i-j|$, we define
\be
b_i = \sum_{j=1}^i a_j ~,~~~ b_{ij} = b_i - b_j
\ee
and try a form
\be
\M_{ij}^{-1} (x) = e^{- b_i x} \left\{ A(x) + B(x) b_{ij} +
C(x) | b_{ij} | \right\} e^{ b_j x}
\ee
The elements with $i,j$ either $1$ or $M$ need special attention
due to the `wrap around' properties of $\M$. (Note that $b_i$ is 
not periodic in its index, but rather $b_{i+M} = b_i +1$.) This
fixes the coefficients $A,B,C$ and the inverse of $\M$ becomes
\be
\M_{ij}^{-1} (x) = e^{- b_{ij} x} \left\{
{1 \over (1-e^x)(1-e^{-x})} +
{b_{ij} (1+e^x ) \over 2(1-e^x)} - {| b_{ij} | \over 2} \right\}
\label{Minv}
\ee
The matrix $\M (0)$
has a zero eigenvalue corresponding to the eigenvector $(1,\dots 1)$.
In order to compute the inverse of $\tM$ it suffices to take the
limit $x \to 0$ in (\ref{Minv}) and discard pieces that are constant
in $i,j$ since they clearly project on the zero-eigenvalue eigenspace.
We find
\be
(\M (x)^{-1})_{ij} = -{1 \over x^2} + {1 \over 12} + 
{b_{ij}^2 - | b_{ij} | \over 2} + O(x) A_{ij}
\ee
Discarding the constant piece and taking $x \to 0$ we 
obtain $\tM^{-1}$. Putting everything together, we finally obtain
\be
H = \half \sum_n p_n^2 
+ \half \sum_{i,j;m \neq n} V_{ij} (\theta_{mn} ) K_{i mn} K_{j nm}
+ \half \sum_{i,j;n} {\tilde V}_{ij} K_{i nn} K_{j nn}
\label{Hpart}
\ee
with the potentials $V_{ij}$ and ${\tilde V}_{ij}$ defined as
\be
V_{ij} (x) = e^{-i b_{ij} x} \left( {1 \over 4 \sin^2 {x \over 2}}
+ i {b_{ij} \over 2} \cot {x \over 2} - {| b_{ij} | \over 2 } \right)
~~,~~~~
{\tilde V}_{ij} = {b_{ij}^2 - | b_{ij} | \over 2}
\label{Vpart}
\ee
This is a generalization of the Sutherland model of particles on
the circle with spin degrees of freedom, encoded by the $K_i$,
where the different $K_i$ couple through a parametric modification
of the inverse-sine-squared potential. 
Interestingly, this is the same classical model as the one obtained 
by Blom and Langmann starting from the 2-dimensional Yang-Mills
point of view \cite{BL}. 
The standard spin-Sutherland
model is recovered either as the limit where all $K_i$ but one
are zero, or as the limit where all $a_i$ but one are zero, in
which case only the sum of all $K_i$ appears in the hamiltonian.

The quantum version of this model proceeds along similar lines.
The constant matrices $P_i$ become, now, $U(N)$ generators
transforming under some fixed representations of $SU(N)$ and 
carrying some $U(1)$ charge (since they are not necessarily traceless).
We shall choose them to be irreducible, the reducible case 
being simply the direct sum of models with one irreducible component
for each $P_i$. Therefore, the model is
labeled by a set of irreps $r_i$ and charges $q_i$ carried by the $P_i$.
The matrices $K_i$ become time-varying $SU(N)$ generators in the same
irrep $r_i$ and with the same charge $q_i$ as $P_i$. In a more
standard notation, putting 
\be
K_i = \sum_{a=1}^{N^2 -1} T^a K_i^a + {q_i \over N}
\ee
with $T^a$ 
the fundamental $SU(N)$ generators, the $K_i^a$ obey the $SU(N)$
algebra for each $i$ and commute for different $i$, while the $q_i$ are
central.
The hamiltonian (\ref{Hpart}) remains valid for the quantum operators
as well; the only term requiring ordering, namely 
$V_{ii} K_{i mn} K_{i nm}$, is automatically symmetrized by the
summation over $m,n$.

To turn the $K_i$ into internal degrees of freedom we follow the
standard construction \cite{MP2} (see also \cite{Jev}). The 
$K_i$ can be realized \`a la Jordan-Wigner in terms of bosonic
oscillators
\be
K_{i mn} = \sum_{k=1}^{d_i} ( A_{ikm}^\dagger A_{ikn} 
- {1 \over N} \delta_{mn} \sum_{l=1}^N A_{ikl}^\dagger A_{ikl} )
+ {q_i \over N} \delta_{mn}
\label{KAA}
\ee
where $d_i$ is the number of rows in the Young tableau of $r_i$. 
The $A_{ikm}$ are a collection of commuting bosonic ladder operators
\be
[ A_{ikm} , A_{jln}^\dagger ] = \delta_{ij} \delta_{kl} \delta_{mn}
\label{PAA}
\ee
The above imbeds $r_i$, as well as all other irreps with up to $d_i$
rows, in the Fock space of $A_{ikm}$.
To simplify the notation and interpretation of the model we will
choose $d_i = 1$ for all $K_i$. We can always achieve the case
$d_i \geq 1$ as a model with
$M' = \sum_{i=1}^M d_i$ matrices with all $d_i =1$. Choosing 
\be
a_j = 0 ~~{\rm for}~~
\sum_{k=1}^{i-1} d_k < j \leq \sum_{k=1}^{i} d_k
\ee
makes all the
$K_j$ in this range, overall $d_i$ in number, appear only 
through their sum in the hamiltonian, reproducing the original
$d_i >1$ matrix. We thus drop the summation and index $k$ in (\ref{KAA}) 
and (\ref{PAA}). 

The gauge constraint (\ref{Gauss}) in terms of (\ref{KAA}) implies
\be
\sum_{i=1}^M A_{in}^\dagger A_{in} = {\rm constant} \equiv \m
~~{\rm for~all}~n, ~~~~
\sum_{i=i}^M q_i = 0
\ee
{}From the above we see that the Fock states generated by the $M$ oscillators
$A_{in}$ for each $n$ transform in the m-fold symmetric irrep of $SU(M)$.
We therefore define the $N$ mutually commuting $SU(M)$ generators
\be
S_{n ij} = A_{in}^\dagger A_{jn} - {\m \over M} \delta_{ij}
\ee
carrying the m-fold symmetric irrep of $SU(M)$. In terms of these,
substituting (\ref{KAA}) in (\ref{Hpart}), we obtain after some algebra
$$
H = \half \sum_n p_n^2 
+ \half \sum_{m \neq n} \left( \sum_{ij} V_{ij} (\theta_{mn} ) 
S_{m ij} S_{n ji} + V_o (\theta_{mn} ) {\m (\m +M) \over M}\right)
$$
\be
+ \half \sum_{n} \sum_{ij} {\tilde V}_{ij} S_{n ii} S_{n jj}
+ {1 \over 2N} \sum_{ij} {\tilde V}_{ij} \left( q_i q_j
- S_{ii} S_{jj} \right)
\label{Hsp}
\ee
where we defined the total spin $S= \sum_{n=1}^N S_n$ and
$V_o (x) \equiv V_{ii} (x)$ = $1/(2 \sin {x \over 2} )^2$.
The above hamiltonian refers
to a system of particles with m-fold symmetric $SU(M)$ 
spins $S_n$. Due, however, to the existence of the matrix $V(x)$
in the coupling of spins, the interactions above are not 
$SU(M)$-invariant (the total spin $S$ is not conserved). Only the
diagonal elements (that is, the Cartan part) of the total spin
are conserved. The appearance of the total spin in (\ref{Hsp})
can be eliminated by choosing the charges
\be
q_i = S_{ii} = {\rm n}_i - {\m N \over M}
\ee
where ${\rm n}_i$ is the number of boxes in $r_i$. (In this construction
the $r_i$ are symmetric, so ${\rm n}_i$ are their lengths.) This is a natural
choice, identifying the $U(1)$ charge to the $Z_N$ charge of $r_i$ and
subtracting the average $Z_N$ charge of all irreps.

In terms of the original matrix problem, the hamiltonian is the
sum of independent laplacians over the space $U(N)$. The $U(1)$ part 
will lead to charges $Q_i$ (the momentum of the `center-of-mass' of each
$U_i$) which, by (\ref{cond}) and (\ref{Uall}), have to satisfy
\be
Q_i - Q_{i+1} = q_i ~~,~~~~ \sum_i a_i Q_i = \sum_n p_n \equiv P
\ee
This fixes the $Q_i$ in terms of the $q_i$ and the total momentum $P$.
Factoring out the $U(1)$ part for each $U_i$, we are left 
with laplacians on the $SU(N)$ manifold. 
It is known that the matrix elements of
each irrep $R$ of $SU(N)$ form a degenerate eigenstate multiplet
of the laplacian with
eigenvalue given by the quadratic Casimir $C_2 (R)$ of the irrep. 
Therefore, the eigenstates of the matrix hamiltonian are
\be
\Psi (\{ U_i \}) = \prod_{i=1}^M (\det U_i )^{Q_i \over N}
R_i ( U_i )_{\alpha_i \beta_i}
\ee
where $R_i$ are irreps of $SU(N)$ ($\alpha_i ,\beta_i$ label their
matrix elements). The energy eigenvalue $E$ and degeneracy $D$
corresponding to this state are
\be
E = \sum_{i=1}^M {a_i \over 2} \left\{ C_2 ( R_i ) + {Q_i^2 \over
N} \right\}~~,~~~~ 
D = \prod_{i=1}^M (dim R_i )^2
\ee
Each
$R_i ( U_i )_{\alpha_i \beta_i}$ transforms under left-rotations 
of $U_i$ in the $R_i$ irrep (acting on the left index) and under
right-rotations of $U_i$ in the conjugate irrep ${\bar R}_i$
(acting on the right index). The condition (\ref{cond}), however,
tells us that the sum of the right-generator for $U_i$ and the
left-generator for $U_{i+1}$ are constrained to be in the $r_i$
irrep of $SU(N)$ carried by $P_i$. Therefore, we must project the
states $\beta_i$ and $\alpha_{i+1}$, transforming in the ${\bar R}_i$
and $R_{i+1}$ respectively, to the subspace of states transforming
in the $r_i$. Call $G({\bar R}_i , \beta_i ; R_{i+1} , 
\alpha_{i+1} | r_i , \gamma_i )$ the Clebsch-Gordan coefficient that 
projects these states to the $\gamma_i$ state of $r_i$. Then the final
energy eigenstate wavefunction for this model becomes
\be
\Psi ( \{ U_i \} ) = \sum_{\{\alpha_j , \beta_j \}} 
\prod_{i=1}^M (\det U_i )^{Q_i \over N}
R_i ( U_i )_{\alpha_i \beta_i} G({\bar R}_i , \beta_i ; 
R_{i+1} , \alpha_{i+1} | r_i , \gamma_i ) 
\ee
The indices $\gamma_i$ do not imply a degeneracy of states
corresponding to the dimensionality of $r_i$, since these are states
fixed by the constraints (\ref{cond}) and are not summed over.
The degeneracy of each state labeled by the $R_i$ is given by the
number of times that the irreps $r_i$ are contained in the direct
products ${\bar R}_i \times R_{i+1}$ or, equivalently, the
number of times that $R_{i+1}$ is contained in $R_i \times
r_i$. Calling this integer $D( R_i , r_i ; R_{i+1} )$, we have
for the total degeneracy
\be
D = \prod_{i=1}^M  D( R_i , r_i ; R_{i+1} )
\ee
Clearly, an irrep $R_i$ can appear in $\Psi$ only if it is
contained in the product $R_i \times r_1 \cdots \times r_N$.
This requires that the total $Z_N$ charge of the $r$'s be zero,
that is, the total number of boxes in the Young tableaux of $r_i$
should be a multiple of $N$. This is indeed the case, the total
number being ${\rm n} = \m N$.
The above, upon proper reinterpretation, gives the spectrum
and wavefunctions in each corresponding sector of the particle-spin 
model as classified by the Cartan elements of the total spin.

We conclude by mentioning that a construction similar to (\ref{KAA}) in 
terms of fermionic oscillators would give rise to a model as in (\ref{Hsp})
but with negative sign (ferromagnetic) spin interactions.

\vspace*{5mm}
I would like to thank E.~Langmann for comments and for communicating
his results \cite{BL} prior to publication.


\begin{thebibliography}{99}

\bibitem{Cal} F.~Calogero, \JMP {\bf 10} (1969) 2191 and 2197; {\bf 12} (1971)
419; \LNC {\bf 13} (1975) 411; F.~Calogero and C.~Marchioro, \LNC
{\bf 13} (1975) 383.

\bibitem{Suth} B.~Sutherland, \PR {\bf A4} (1971) 2019; {\bf 5} (1972) 1372;
\PRL {\bf 34} (1975) 1083.

\bibitem{Mos} J.~Moser, {\it Adv. Math.} {\bf 16} (1975) 1.

\bibitem{GH} J.~Gibbons and T.~Hermsen, {\it Physica} {\bf D11} (1984) 337.

\bibitem{Woj} S.~Wojciechowski, \PL {\bf A111} (1985) 101.

\bibitem{HH} Z.N.C.~Ha and F.D.M.~Haldane, \PR {\bf B46} (1992) 9359.

\bibitem{Kaw} N.~Kawakami, \PR {\bf B46} (1992) 1005 and 3191.

\bibitem{MP1} J.A.~Minahan and A.P.~Polychronakos, \PL {\bf B302} (1993) 265.

\bibitem{HW} K.~Hikami and M.~Wadati, \PL {\bf A173} (1993) 263.

\bibitem{Hal} F.D.M.~Haldane, \PRL {\bf 60} (1988) 635;
{\bf 66} (1991) 1529.

\bibitem{Sha} B.S.~Shastry, \PRL {\bf 60} (1988) 639; {\bf 69}
(1992) 164.

\bibitem{FM} M.~Fowler and J.A.~Minahan, \PRL {\bf 70} (1993) 2325.

\bibitem{Scal} A.P.~Polychronakos, \PRL {\bf 70} (1993) 2329;
\NP {\bf B419} (1994) 553.

\bibitem{GN} A.~Gorsky and N.~Nekrasov, \NP {\bf B414} (1994) 213.

\bibitem{MP2} J.A.~Minahan and A.P.~Polychronakos, \PL {\bf B326}
(1994) 288.

\bibitem{LSK} E.~Langmann, M.~Salmhofer and A.~Kovner, \MPL 
{\bf A9} (1994) 2913.
 
\bibitem{KKS} D.~Kazhdan, B.~Kostant and S.~Sternberg, 
{\it Comm. Pure Appl. Math.} {\bf 31} (1978) 481.

\bibitem{OP} M.A.~Olshanetsky and A.M.~Perelomov, {\it Phys. Rep.} {\bf 71} 
(1981) 314; {\bf 94} (1983) 6.

\bibitem{BL} J.~Blom and E.~Langmann, solv-int/9804007.

\bibitem{Jev} J.~Avan and A.~Jevicki, \NP {\bf B469} (1996) 287.


\end{thebibliography}
\end{document}